\documentclass[]{revtex4}
\usepackage{graphicx}
\usepackage{fancyhdr}
\begin{document}

\title{ {\bf QCD sum rules analysis  of  the $B_{s}\rightarrow D_{sJ}(2460)l\nu $ decay   }}
\author{T. M. Aliev, K. Azizi *, A. Ozpineci\\ Physics Department, Middle
East Technical University, Ankara, Turkey\\**: Speaker}
\fancyhead[C]{\it {International Conference on Hadron Physics, 30
Aug - 3 Sep 2007, Canakkale, Turkey}} \fancyhead[RO,LE]{\thepage}
\setlength{\baselineskip}{24pt} \maketitle
\setlength{\baselineskip}{7mm}
 Using three point QCD sum rules method, the form factors relevant to the semileptonic
 $B_{s}\rightarrow D_{sJ}(2460)\ell\nu$ decay are calculated.
The $q^2$ dependencies of these form factors
 are evaluated. The dependence of the asymmetry parameter
 $\alpha$, characterizing the polarization of $D_{sJ}$ meson, on $q^2$ is studied. This study gives useful information about
 the structure of the $D_{sJ}$ meson.  Finally the branching ratio
 of this decay is
 also estimated and is shown that it can be easily detected at LHC.
 Note that this talk is based on the original work in \cite{0}.

\maketitle

\thispagestyle{fancy}


%In the present paper
\section{Introduction}

 Recently, very exciting experimental results have been obtained in charmed hadron spectroscopy. The observation
 of two narrow resonances with charm and strangeness, $D_{sJ}(2317)$ in the $D_{s}\pi^{0}$ invariant mass distribution
\cite{1,2,3,4,5,6}, and $D_{sJ}(2460)$ in the $D_{s}^{\ast}\pi^{0}$
and $D_{s}\gamma$ mass distribution \cite{2,3,4,6,7,8}, has raised
discussions about the nature of these states and their quark content
\cite{9,10}.
  Analysis of the $D_{s_{0}}(2317)\rightarrow
D_{s}^{\ast}\gamma$, $D_{sJ}(2460)\rightarrow
    D_{s}^{\ast}\gamma$ and $ D_{S_{J}}(2460)\rightarrow D_{s_{0}}(2317)\gamma$ indicates that the quark
    content of these mesons are probably $\overline{c}s$ \cite{11}. In \cite{11} it
    is also shown that finite quark mass effects for the $c$-quark give non-negligible corrections.

    When LHC begins operation, an abundant number of $B_{s}$ mesons will
    be produced creating a real possibility for studying the
    properties of $B_{s}$ meson and its various decay
    channels. One of the possible decay channels  of $B_{s}$ meson
    is its
    semileptonic $B_{s}\rightarrow D_{sJ}(2460)l\nu$ decay.
    Analysis of this decay might yield useful information for understanding the
    structure of the $ D_{sJ}(2460)$ meson.

    It is well known that the
    semileptonic decays of heavy flavored mesons are very promising tools
    for the determination of the elements of the CKM
    matrix, leptonic decay constants as well as the origin of the CP
    violation. In semileptonic decays the long distance dynamics are
    parameterized by transition form factors, calculation of which is
    a central problem for these decays.

    Obviously, for the calculation of the transition form factors,
    nonperturbative approaches are needed. Among the nonperturbative
    approaches, the QCD sum rules method \cite{12} received special
    attention, because this method is based on the fundamental QCD
    Lagrangian. This method has been successfully applied to a wide
    variety of problems in hadron physics(for a review see \cite{13}). The
    semileptonic decay $D\rightarrow \overline{K}^{0}l\nu$ is
    studied using the QCD sum rules with three point correlation
    function in \cite{14}. Then, the semileptonic decays $D^{+}\rightarrow
    K^{0\ast}e^{+}\nu~e$ \cite{15}, $D\rightarrow \pi~l\nu$ \cite{16}, $D\rightarrow
    \rho~l\overline{\nu}e$ \cite{17} and $B\rightarrow
    D(D^{\ast})l\nu~e$ \cite{18} are studied in the same framework.
    In the
    present work we study the semileptonic decay of $B_{s}$ meson
    to positive parity $D_{sJ}(2460)$ meson,i.e, $B_{s}\rightarrow
    D_{sJ}(2460)\ell\nu$, within QCD sum rules method. Note that, in \cite{19},
    the decay $B_{s}\rightarrow D_{s_{0}}(2317)l\nu$ has been studied using the QCD sum rules.

    The paper is organized
    as follows: In section  II  the sum rules for the transition form factors are
    calculated; section  III  is devoted to the numerical analysis, discussion and our conclusions.

%%%
%%%
\section{Sum rules for the $B_{s}\rightarrow D_{sJ}(2460)\ell\nu$ transition form factors }
The $B_s \rightarrow  D_{sJ}$ transition proceeds by the
$b\rightarrow c$ transition at the quark level. The matrix element
for the quark level process can be written as:
\begin{equation}\label{lelement}
M_{q}=\frac{G_{F}}{\sqrt{2}} V_{cb}~\overline{\nu}
~\gamma_{\mu}(1-\gamma_{5})l~\overline{c}
~\gamma_{\mu}(1-\gamma_{5}) b \\
\end{equation}
In order to obtain the matrix elements for $B_{s}\rightarrow
D_{sJ}(2460)\ell\nu$ decay, we need to sandwich Eq. (\ref{lelement})
between initial and final meson states. So, the amplitude of the
$B_{s}\rightarrow D_{sJ}(2460)\ell\nu$ decay can be written as:
\begin{equation}\label{2au}
M=\frac{G_{F}}{\sqrt{2}} V_{cb}~\overline{\nu}
~\gamma_{\mu}(1-\gamma_{5})l<D_{sJ}\mid~\overline{c}
~\gamma_{\mu}(1-\gamma_{5}) b\mid B_{s}>
\end{equation}
 The main problem is the calculation of the matrix element
$<D_{sJ}\mid\overline{c}\gamma_{\mu}(1-\gamma_{5}) b\mid B_{s}>$
appearing in Eq. (\ref{2au}). Both vector and axial vector part of
  $~\overline{c}~\gamma_{\mu}(1-\gamma_{5}) b~$  contribute to the
matrix element considered above. From Lorentz invariance and parity
considerations, this matrix element can be parameterized in terms of
the form factors in the following way:
\begin{equation}\label{3au}
<D_{sJ}(p',\varepsilon)\mid\overline{c}\gamma_{\mu}\gamma_{5} b\mid
B_s(p)>=\frac{f_{V}(q^2)}{(m_{B_{s}}+m_{D_{sJ}})}\varepsilon_{\mu\nu\alpha\beta}
\varepsilon^{\ast\nu}p^\alpha p'^\beta
\end{equation}
\begin{eqnarray}\label{4au}
< D_{sJ}(p',\varepsilon)\mid\overline{c}\gamma_{\mu} b\mid B_{s}(p)>
&=&i\left[f_{0}(q^2)(m_{B_{s}} +m_{D_{sJ}})\varepsilon_{\mu}^{\ast}
\right. \nonumber \\
+
\frac{f_{+}(q^2)}{(m_{B_{s}}+m_{D_{sJ}})}(\varepsilon^{\ast}p)P_{\mu}
&+& \left.
\frac{f_-(q^2)}{(m_{B_{s}}+m_{D_{sJ}})}(\varepsilon^{\ast}p)q_{\mu}\right]
\end{eqnarray}
where $f_{V}(q^2)$, $f_{0}(q^2)$, $f_{+}(q^2)$ and $f_{-}(q^2)$ are
the transition form factors and $P_{\mu}=(p+p')_{\mu}$,
$q_{\mu}=(p-p')_{\mu}$. In all following discussions, for customary,
we will use following redefinitions:
\begin{eqnarray}\label{eq5}
f_{V}'(q^2)&=&\frac{f_{V}(q^2)}{(m_{B_{s}}+m_{D_{sJ}})}~,~~~~~~~~~~~~f_{0}'(q^2)=f_{0}(q^2)(m_{B_{s}}
+m_{D_{sJ}})\nonumber
\\
f_{+}'(q^2)&=&\frac{f_{+}(q^2)}{(m_{B_{s}}+m_{D_{sJ}})}~,~~~~~~~~~~~~
f_{-}'(q^2)=\frac{f_{- }(q^2)}{(m_{B_{s}}+m_{D_{sJ}})}
\end{eqnarray}
For the calculation of these  form factors,  QCD sum rules method
will be employed. We start by considering the following correlator:
\begin{equation}\label{6au}
\Pi _{\mu\nu}^{V;A}(p^2,p'^2,q^2)=i^2\int
d^{4}xd^4ye^{-ipx}e^{ip'y}<0\mid T[J _{\nu D_{sJ}}(y)
J_{\mu}^{V;A}(0) J_{B_{s}}(x)]\mid  0>
\end{equation}
where $J _{\nu D_{sJ}}(y)=\overline{s}\gamma_{\nu} \gamma_{5}c$,
$J_{B_{s}}(x)=\overline{b}\gamma_{5}s$ ,
 $J_{\mu}^{V}=~\overline{c}\gamma_{\mu}b $ and $J_{\mu}^{A}=~\overline{c}\gamma_{\mu}\gamma_{5}b$
 are the interpolating currents of the  $D_{sJ}$, $B_{s} $,
 vector and axial vector
currents respectively.

To calculate the phenomenological part of the correlator given in
Eq. (\ref{6au}), two complete sets of intermediate states with the
same quantum number as the currents $J_{D_{sJ}}$ and $J_{B_{s}}$
respectively are inserted. As a result of this procedure we get the
following representation of the above-mentioned correlator:
\begin{eqnarray} \label{7au}
&&\Pi _{\mu\nu}^{V,A}(p^2,p'^2,q^2)=
\nonumber \\
&& \frac{<0\mid J_{D_{sJ}}^{\nu} \mid
D_{sJ}(p')^{\varepsilon}><D_{sJ}(p')^{\varepsilon}\mid
J_{\mu}^{V,A}\mid B_{s}(p)><B_{s}(p)\mid J_{Bs}\mid
0>}{(p'^2-m_{D_{sJ}}^2)(p^2-m_{Bs}^2)}+\cdots
\nonumber \\
\end{eqnarray}
 where $\cdots$ represent contributions coming from higher states and continuum. The matrix
 elements in Eq. (\ref{7au}) are defined in the standard way as:
\begin{equation}\label{8au}
 <0\mid J^{\nu}_{D_{sJ}} \mid
D_{sJ}(p')>=f_{D_{sJ}}m_{D_{sJ}}\varepsilon^{\nu}~,~~<B_{s}(p)\mid
J_{Bs}\mid 0>=-i\frac{f_{B_{s}}m_{B_{s}}^2}{m_{b}+m_{s}}
\end{equation}
where $f_{D_{sJ}}$ and $f_{B_{s}}$  are the leptonic decay constants
of $D_{sJ} $ and $B_{s}$ mesons, respectively. Using Eq.
(\ref{3au}), Eq. (\ref{4au}) and Eq. (\ref{8au}) and performing
summation over the polarization of the $D_{sJ}$ meson, Eq.
(\ref{7au}) can be written as:
\begin{eqnarray}\label{9amplitude}
\Pi_{\mu\nu}^{V}(p^2,p'^2,q^2)&=&-\frac{f_{B_{s}}m_{B_{s}}^2}{(m_{b}+m_{s})}\frac{f_{D_{sJ}}m_{D_{sJ}}}
{(p'^2-m_{D_{sJ}}^2)(p^2-m_{Bs}^2)} \times
[f_{0}'g_{\mu\nu}+f_{+}'P_{\mu}p_{\nu} \nonumber
\\ &+&f_{-}'q_{\mu}p_{\nu}]+
\mbox{excited states.}\nonumber
\\\Pi_{\mu\nu}^{A}(p^2,p'^2,q^2)&=&
-i\varepsilon_{\mu\nu\alpha\beta}p'^{\alpha}p^{\beta}\frac{f_{B_{s}}m_{B_{s}}^2}{(m_{b}+m_{s})}\frac{f_{D_{sJ}}m_{D_{sJ}}}
{(p'^2-m_{D_{sJ}}^2)(p^2-m_{Bs}^2)}f_{V}' +
\nonumber \\
&&\mbox{excited states.}
\end{eqnarray}

In accordance with the QCD sum rules philosophy, $\Pi
_{\mu\nu}(p^2,p'^2,q^2)$ can also be calculated from QCD side with
the help of the operator product expansion(OPE) in the deep
Euclidean region  $p^2 \ll (m_{b}+m_{c})^2 $ and $p'^2 \ll
(m_{c}+m_{s})^2$.
%To obtain the sum rules for the form factors, the two different representations of
%$\Pi _{\mu\nu}(p^2,p'^2,q^2)$ are equated using spectral representation.
The theoretical part of the correlator is calculated by means of
OPE, and up to operators having dimension $d=6$, it is determined by
the bare-loop and the power corrections from the operators with
$d=3$, $<\overline{\psi}\psi>$, $d=4$, $m_{s}<\overline{\psi}\psi>$,
$d=5$, $m_{0}^{2}<\overline{\psi}\psi>$ and $d=6$,
$<\overline{\psi}\psi\bar \psi \psi>$. In calculating the $d=6$
operator, vacuum saturation approximation is used to set
$<\overline{\psi}\psi\bar \psi \psi> = <\overline{\psi}\psi>^2$.
 In calculating the
bare-loop contribution, we first write the double dispersion
representation for the coefficients of corresponding Lorentz
structures appearing in the correlation function as:
\begin{equation}\label{10au}
\Pi_i^{'per}=-\frac{1}{(2\pi)^2}\int
dsds'\frac{\rho_{i}(s,s',q^2)}{(s-p^2)(s'-p'^2)}+\textrm{
subtraction terms}
\end{equation}
The spectral densities $\rho_{i}(s,s',q^2)$ can be calculated from
the usual Feynman integral with the help of Cutkosky rules, i.e. by
replacing the quark propagators with Dirac delta functions:
$\frac{1}{p^2-m^2}\rightarrow-2\pi\delta(p^2-m^2),$ which implies
that all quarks are real. After standard calculations for the
corresponding spectral densities we obtain:
\begin{eqnarray}\label{11au}
\rho_{V}(s,s',q^2)&=&N_{c}I_{0}(s,s',q^2)\left[{m_{s}+(m_{s}-m_{b})B_{1}+(m_{s}+m_{c})B_{2}}\right],\nonumber\\
\rho_{0}(s,s',q^2)&=&N_{c}I_{0}(s,s',q^2)[8(m_{b}-m_{s})A_{1}-4m_{b}m_{c}m_{s}\nonumber\\&+&
4(m_{s}- m_{b}+m_{c
})m_{s}^2-2(m_{s}+m_{c})(\Delta+m_{s}^2)\nonumber
\\&-&
2(m_{s}-m_{b})(\Delta'+m_{s}^2)+2m_{s}u]\nonumber \\
\rho_{+}(s,s',q^2)&=&N_{c}I_{0}(s,s',q^2)[4(m_{b}-m_{s})(A_{2}+A_{3})+2(m_{b}-3m_{s})B_{1}
\nonumber \\
&& -2(m_{c}+m_{s})B_{2}-2m_{s}]
,\nonumber \\
\rho_{-}(s,s',q^2)&=&N_{c}I_{0}(s,s',q^2)[4(m_{b}-m_{s})(A_{2}-A_{3})-2(m_{b}+m_{s})B_{1}
\nonumber \\
&& +2(m_{c}+m_{s})B_{2} +2m_{s}]\nonumber \\
\end{eqnarray}
where
\begin{eqnarray}\label{12}
I_{0}(s,s',q^2)&=&\frac{1}{4\lambda^{1/2}(s,s',q^2)},\nonumber\\
 \lambda(s,s',q^2)&=&s^2+s'^2+q^4-2sq^2-2s'q^2-2ss',\nonumber \\
\Delta'&=&(s'-m_{c}^2 + m_{s}^2),\nonumber\\
\Delta&= &(s-m_{b}^2 +
m_{s}^2),\nonumber\\
 u &=& s + s' - q^2,\nonumber\\
 B_{1}&=&\frac{1}{\lambda(s,s',q^2)}[2s'\Delta-\Delta'u],\nonumber\\
 B_{2}&=&\frac{1}{\lambda(s,s',q^2)}[2s\Delta'-\Delta u],\nonumber\\
 A_{1}&=&\frac{1}{2\lambda(s,s',q^2)}[\Delta'^{2}s+2\Delta'm_{s}^2s+m_{s}^4s+\Delta^2s'+2 \Delta m_{s}^2s'
 \nonumber \\
 && +m_{s}^4s'- 4m_{s}^2ss'-\Delta\Delta'u-\Delta m_{s}^2u-\Delta'm_{s}^2u-m_{s}^4u+m_{s}^2u^2],\nonumber\\
 A_{2}&=&\frac{1}{\lambda^{2}(s,s',q^2)}[2\Delta'^2ss'+4\Delta'm_{s}^2ss'+2m_{s}^4ss'+6\Delta^2s'^2
 \nonumber \\
 && +12\Delta m_{s}^2s'^2 +6m_{s}^4s'^2-8m_{s}^2ss'^2-6\Delta\Delta's'u
 \nonumber \\
 && -6\Delta m_{s}^2s'u-6\Delta'm_{s}^2s'u-6m_{s}^4s'u+\Delta'^2u^2+2\Delta'
 m_{s}^2u^2
 \nonumber\\
 &&+m_{s}^4u^2+2m_{s}^2s'u^2],\nonumber\\
 A_{3}&=&\frac{1}{\lambda^{2}(s,s',q^2)}[4\Delta\Delta'ss'+4\Delta m_{s}^2ss'+4\Delta'm_{s}^2ss'+4m_{s}^4ss'
 \nonumber \\
 &&-3\Delta'^2su- 6\Delta'm_{s}^2su -3m_{s}^4su-3\Delta^2s'u-6\Delta m_{s}^2s'u
 \nonumber \\
 &&-3m_{s}^4s'u+4m_{s}^2ss'u+2\Delta\Delta'u^2+
 2\Delta m_{s}^2u^2+2\Delta' m_{s}^2u^2
 \nonumber \\
 &&+2m_{s}^4u^2-m_{s}^2u^3]\nonumber\\
 \end{eqnarray}
 The subscripts V, 0 and $\pm$ correspond to the coefficients of the
 structures proportional to $i\varepsilon_{\mu\nu\alpha\beta}p'^{\alpha}p^{\beta}$, $g_{\mu\nu}$ and $\frac{1}{2}(p_{\mu}p_{\nu}
 \pm p'_{\mu}p_{\nu})$ respectively. In Eq. (\ref{11au}) $N_{c}=3$ is the number of colors.

 The integration region for the perturbative contribution
 in Eq. (\ref{10au}) is determined from the condition that arguments of the
 three $\delta$ functions must vanish simultaneously. The physical
 region in s and s' plane is described by the following
 inequalities:\\
 \begin{equation}\label{13au}
 -1\leq\frac{2ss'+(s+s'-q^2)(m_{b}^2-s-m_{s}^2)+(m_{s}^2-m_{c}^2)2s}{\lambda^{1/2}(m_{b}^2,s,m_{s}^2)\lambda^{1/2}(s,s',q^2)}\leq+1
\end{equation}

For the contribution of power corrections, i.e. the contributions of
operators with dimensions $d=3$, $4$ and $5$, we obtain the
following results:
\begin{eqnarray}\label{14au}
f_{V}^{'(3)}+f_{V}^{'(4)}+f_{V}^{'(5)}&=&\frac{1}{rr'}<\overline{s}s>-\frac{m_{s}}{2}<\overline{s}s>[\frac{-m_{c}}{rr'^2}+\frac{m_{b}}{r'r^2}]
\nonumber \\ && +\frac{m_{s}^2}{2} <\overline{s}s>
[\frac{2m_{c}^2}{r'^3r}+\frac{m_{b}^2+m_{c}^2-q^2}{r'^2r^2}+\frac{2m_{b}^2}{r'r^3}]
\nonumber \\
&& -\frac{m_{0}^2}{6}<\overline{s}s>
[\frac{3m_{c}^2}{r'^3r}+\frac{3m_{b}^2}{r'r^3}+\frac{2}{r'r^2}i
\nonumber \\ && +\frac{2m_{b}^2+2m_{c}^2+m_{b}m_{c} -2q^2}{r'^2r^2}]
\nonumber \\
f_{0}^{'(3)}+f_{0}^{'(4)}+f_{0}^{'(5)}&=&\frac{(m_{b}-m_{c})^2-q^2}{2rr'}<\overline{s}s>
\nonumber \\ && +\frac{m_{s}}{4}
<\overline{s}s>[\frac{-2m_{b}m_{c}^2+m_{c}m_{b}^{2}+m_{c}^3-m_{c}q^2}{rr'^2}\nonumber\\&&-\frac{m_{c}+m_{b}}{rr'}
+\frac{2m_{c}m_{b}^2-m_{b}^3-m_{b}m_{c}^2+m_{b}q ^2}{r'r^2}]\nonumber \\
&&+\frac{m_{s}^2}{16}<\overline{s}s>\left\{\frac{-16m_{b}m_{c}^3+8m_{c}^2m_{b}^2+8m_{c}^4-8m_{c}^2q^2}{r'^3r}
\right.
\nonumber\\
&&+\frac{-16m_{b}^3m_{c}
+8m_{b}^4+8m_{c}^2m_{b}^2-8m_{b}^2q^2}{r'r^3}\nonumber \\
&&+\frac{4m_{c}^2-8m_{b}m_{c}+4m_{b}^{2}-4q^2}{r'^2r}
\nonumber \\
&& +\frac{4m_{c}^2-8m_{b}m_{c}+4m_{b}^{2}-4q^2}{r'r^2}-\frac{8}{r'r}
 \nonumber
\\&&+\frac{1}{r'^2 r^2} \left[-8m_{b}^3m_{c}-8m_{b}m_{c}^3+8m_{b}m_{c}q^2+4m_{b}^{4} \right.
\nonumber \\ && +\left. \left. 8m_{c}^2m_{b}^2
+4m_{c}^4-8m_{b}^2q^2-8m_{c}^2q^2+4q^4\right]\right\}
\nonumber \\
&&-\frac{m_{0}^2}{12}<\overline{s}s>[\frac{3m_{c}^2(m_{c}^2+m_{b}^2-2m_{b}m_{c}-q^2)}{r'^3r}
\nonumber \\
&&+ 3 m_b^2\frac{m_{c}^2 +m_{b}^2-2m_{b}m_{c}-q^2}{r'r^3}\nonumber
\\&&+
\frac{-3m_{b}m_{c}(m_{c}^2+m_{b}^2-q^2)+2(m_{c}^2+m_{b}^2-q^2)^2-2m_{c}^2m_{b}^2}{r'^2r^2}\nonumber\\&&+\frac{3m_{c}
(m_{c}-m_{b}) +2(m_{b}^2-q^2)}{rr'^2}
\nonumber \\
&& +\frac{3m_{b}(-3m_{c}+m_{b})+4(m_{c}^2-q^2)}{r^2r'}-\frac{2}{rr'}]\nonumber \\
f_{+}^{'(3)}+f_{+}^{'(4)}+f_{+}^{'(5)}&=&-\frac{1}{2rr'}<\overline{s}s>+\frac{m_{s}}{4}<\overline{s}s>[\frac{-m_{c}}{rr'^2}+\frac{m_{b}}{r'r^2}]
\nonumber \\
&&+\frac{m_{s}^2}{32} <\overline{s}s>
[-\frac{16m_{c}^2}{r'^3r}-\frac{16m_{b}^2}{r'r^3}
+\frac{16}{r'r^2}+\frac{-8m_{b}^2-8m_{c}^2+8q^2}{r^2r'^2}] \nonumber
\\ &&
+\frac{m_{0}^2}{12}<\overline{s}s>[\frac{3m_{c}^2}{r'^3r}+\frac{3m_{b}^2}{r'r^3}-
\frac{2}{r'r^2}\nonumber \\&&+\frac{2m_{b}^2+2m_{c}^2+m_{b}m_{c}-2q^2}{r'^2r^2}]\nonumber \\
f_{-}^{'(3)}+f_{-}^{'(4)}+f_{-}^{'(5)}&=&\frac{1}{2rr'}<\overline{s}s>-\frac{m_{s}}{4}<\overline{s}s>[\frac{-m_{c}}{rr'^2}+\frac{m_{b}}{r'r^2}]
\nonumber \\ && +\frac{m_{s}^2}{32} <\overline{s}s>
[\frac{16m_{c}^2}{r'^3r}+\frac{16m_{b}^2}{r'r^3}
+\frac{16}{r'r^2}+\frac{8m_{b}^2+8m_{c}^2-8q^2}{r^2r'^2}] \nonumber
\\ &&
-\frac{m_{0}^2}{12}<\overline{s}s>[\frac{3m_{c}^2}{r'^3r}+\frac{3m_{b}^2}{r'r^3}+
\frac{6}{r'r^2}\nonumber
\\&&+\frac{2m_{b}^2+2m_{c}^2+m_{b}m_{c}-2q^2}{r'^2r^2}]
\end{eqnarray}
where $r=p^2-m_{b}^2,r'=p'^2-m_{c}^2$. We would like to note that
the contributions of operators with $d=6$ are also calculated.
Numerically their contributions to the corresponding sum rules
turned out to be very small and therefore we did not present their
explicit expressions. Note also that, in the present work we neglect
the $\alpha_{s}$ corrections to the bare loop. For consistency, we
also neglect $\alpha_{s}$ corrections in determination of the
leptonic decay constants $f_{B_{s}}$ and $f_{D_{sJ}}$.

The QCD sum rules for the form factors $f'_{V}$, $f'_{0}$, $f'_{+}$
and $f'_{-}$ is obtained by equating the phenomenological expression
given in Eq. (\ref{9amplitude}) and the OPE expression given by Eqs.
(\ref{11au}-\ref{14au}) and applying double Borel transformations
with respect to the variables $p^2$ and $p'^2$ ($p^2\rightarrow
M_{1}^2,p'^2\rightarrow M_{2}^2$) in order to suppress the
contributions of higher states and continuum:
\begin{eqnarray}\label{15au}
f'_{i}(q^2)=-\frac{(m_{b}+m_{s})
}{f_{B_{s}}m_{B_{s}}^2}\frac{1}{f_{D_{sJ}}m_{D_{sJ}}}e^{m_{B_{s}}^2/M_{1}^2+m_{D_{sJ}}^2/M_{2}^2}
\nonumber
\\\times[-\frac{1}{(2\pi)^2)}\int_{(m_b+m_s)^2}^{s_0} ds \int_{(m_c+m_s)^2}^{s_0'} ds'\rho_{i}(s,s',q^2)e^{-s/M_{1}^2-s'/M_{2}^2}\nonumber
\\+\hat{B}(f_{i}^{(3)}+f_{i}^{(4)}+f_{i}^{(5)})]\nonumber\\
\end{eqnarray}
where $i=V,0$ and $\pm$, and $\hat B$ denotes the double Borel
transformation operator. In Eq. (\ref{15au}), in order to subtract
the contributions of the higher states and the continuum,
quark-hadron duality assumption is used, i.e. it is assumed that
\begin{eqnarray}
\rho^{higher states}(s,s') = \rho^{OPE}(s,s') \theta(s-s_0)
\theta(s-s'_0)
\end{eqnarray}
In calculations the following rule for double Borel
transformations is used:\\
\begin{equation}\label{16au}
\hat{B}\frac{1}{r^m}\frac{1}{r'^n}\rightarrow(-1)^{m+n}\frac{1}{\Gamma(m)}\frac{1}{\Gamma
(n)}e^{m_{b}^2/M^{2}}e^{m_{c}^2/M'^{2}}\frac{1}{(M^{2})^{m-1}(M'^{2})^{n-1}}.
\end{equation}
%
%%%
%%%
\section{Numerical analysis}
In this section we present our numerical analysis for the form
factors $f_{V}(q^2)$, $f_{0}(q^2)$, $f_{+}(q^2)$ and $f_{-}(q^2)$.
From sum rule expressions of these form factors we see that  the
condensates, leptonic decay constants of $B_{s}$ and $D_{sJ}$
mesons, continuum thresholds $s_{0}$ and  $s'_{0} $ and Borel
parameters $M_{1}^2$ and $M_{2}^2$ are the main input parameters. In
further numerical analysis we choose the value of the condensates at
a fixed renormalization scale of about $1$ GeV. The values of the
condensates are\cite{20} :
$<\overline{\psi}\psi\mid_{\mu=1~GeV}>=-(240\pm10~MeV)^3$,
$<\overline{s}s>=(0.8\pm0.2)<\overline{\psi}\psi>$ and
$m_{0}^2=0.8~GeV^2$.
 The quark masses are taken to be $ m_{c}(\mu=m_{c})=
 1.275\pm
 0.015~ GeV$, $m_{s}(1~ GeV)\simeq 142 ~MeV$ \cite{21} and $m_{b} =
(4.7\pm
 0.1)~GeV$ \cite{20} also the mesons masses are taken to be $m_{D_{sJ}}=2.46~GeV$ and $ m_{B_{s}}=5.3~GeV$. For
 the values of the leptonic decay
constants of $B_{s}$ and $D_{sJ} $ mesons we use the results
obtained from two-point QCD analysis: $f_{B_{s}} = 209\pm
 38~ MeV $ \cite{13} and $f_{D_{sJ}} =225\pm25
  ~MeV $\cite{11}. The threshold parameters
$s_{0}$ and $s_{0}' $ are also determined from the two-point QCD sum
rules: $s_{0} =(35\pm 2)~ GeV^2$ \cite{12} and $s_{0}' =9~ GeV^2 $
\cite{11}. The Borel parameters $M_{1}^2$ and $M_{2}^2 $ are
auxiliary quantities and therefore the results of physical
quantities should not depend on them. In QCD sum rule method, OPE is
truncated at some finite order, leaving a residual dependence on the
Borel parameters. For this reason, working regions for the Borel
parameters should be chosen such that in these regions  form factors
are practically independent of them. The working regions for the
Borel parameters $M_{1}^2 $ and $M_{2}^2$ can be determined by
requiring that, on the one side, the continuum contribution should
be small, and on the other side, the contribution of the operator
with the highest dimension should be small. As a result of the
above-mentioned requirements, the working regions are determined to
be $ 10~ GeV^2 < M_{1}^2 <20~ GeV^2 $ and $ 4~ GeV^2 <M_{2}^2 <10
~GeV^2$.

\begin{figure}
\vspace*{0cm}
\begin{center}
\includegraphics[width=11cm]{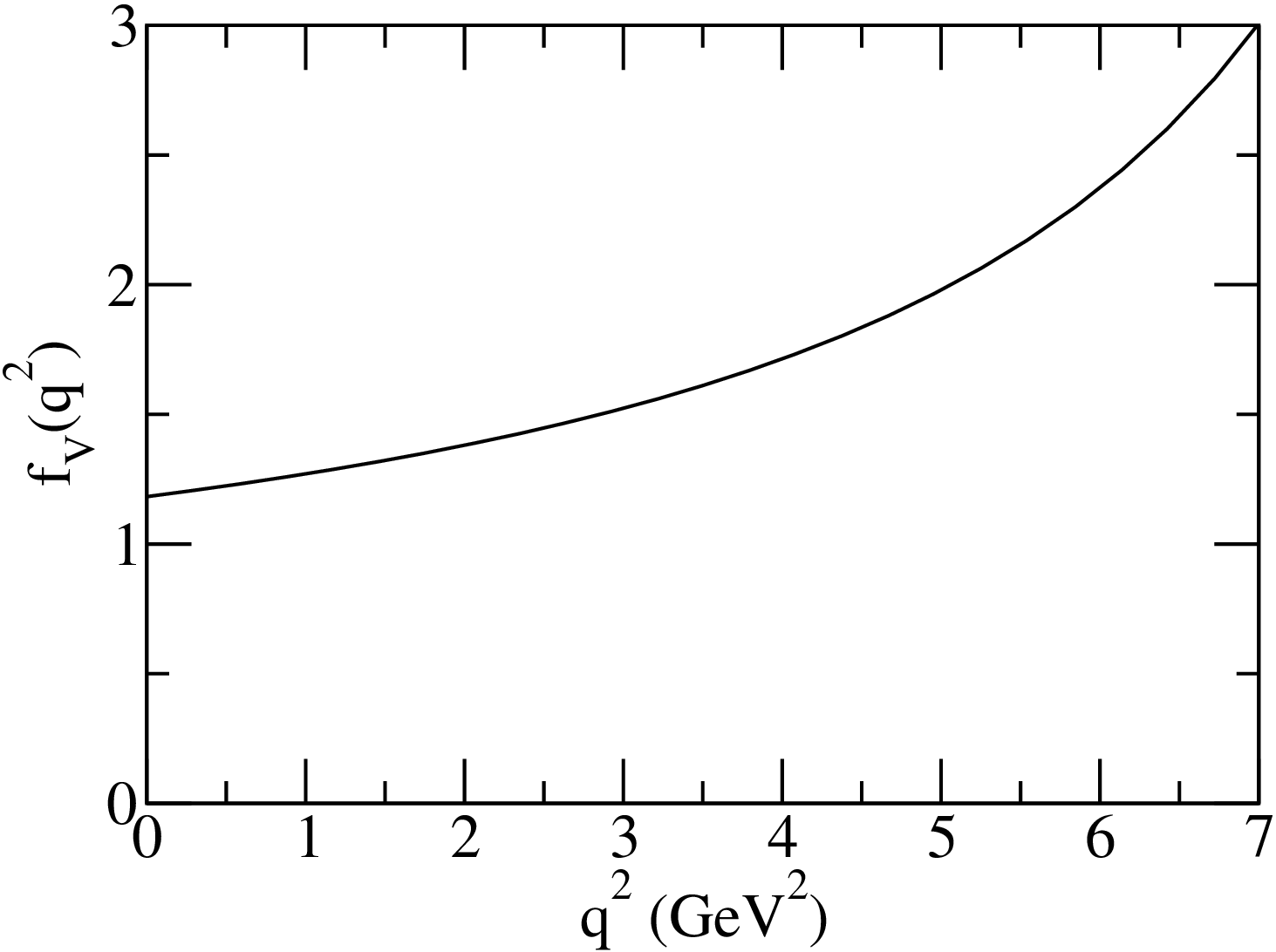}
\end{center}
\caption{The dependence of $f_{V}$ on
 $q^2$ at $M_{1}^2=15~GeV^2$, $M_{2}^2=6~GeV^2$, $s_{0}=35~GeV^2$ and $s_{0}'=9~GeV^2$. } \label{fig1}
\end{figure}

\begin{figure}
\begin{center}
\includegraphics[width=10cm]{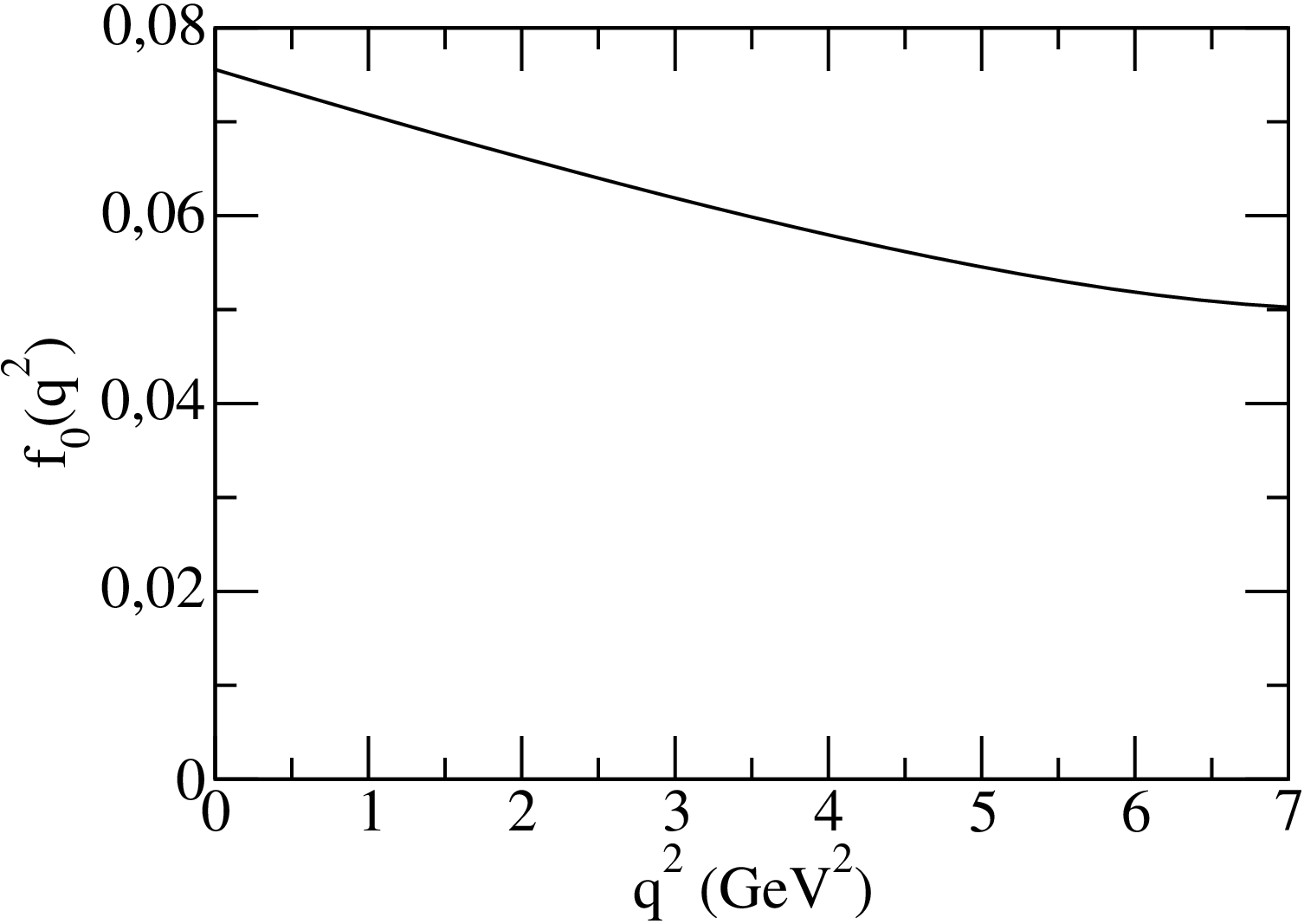}
\end{center}
\caption{ The dependence of $f_{0}$ on
 $q^2$ at $M_{1}^2=15~GeV^2$, $M_{2}^2=6~GeV^2$, $s_{0}=35~GeV^2$ and $s_{0}'=9~GeV^2$.} \label{fig2}
\end{figure}

\begin{figure}
\vspace{0cm}
\begin{center}
\includegraphics[width=10cm]{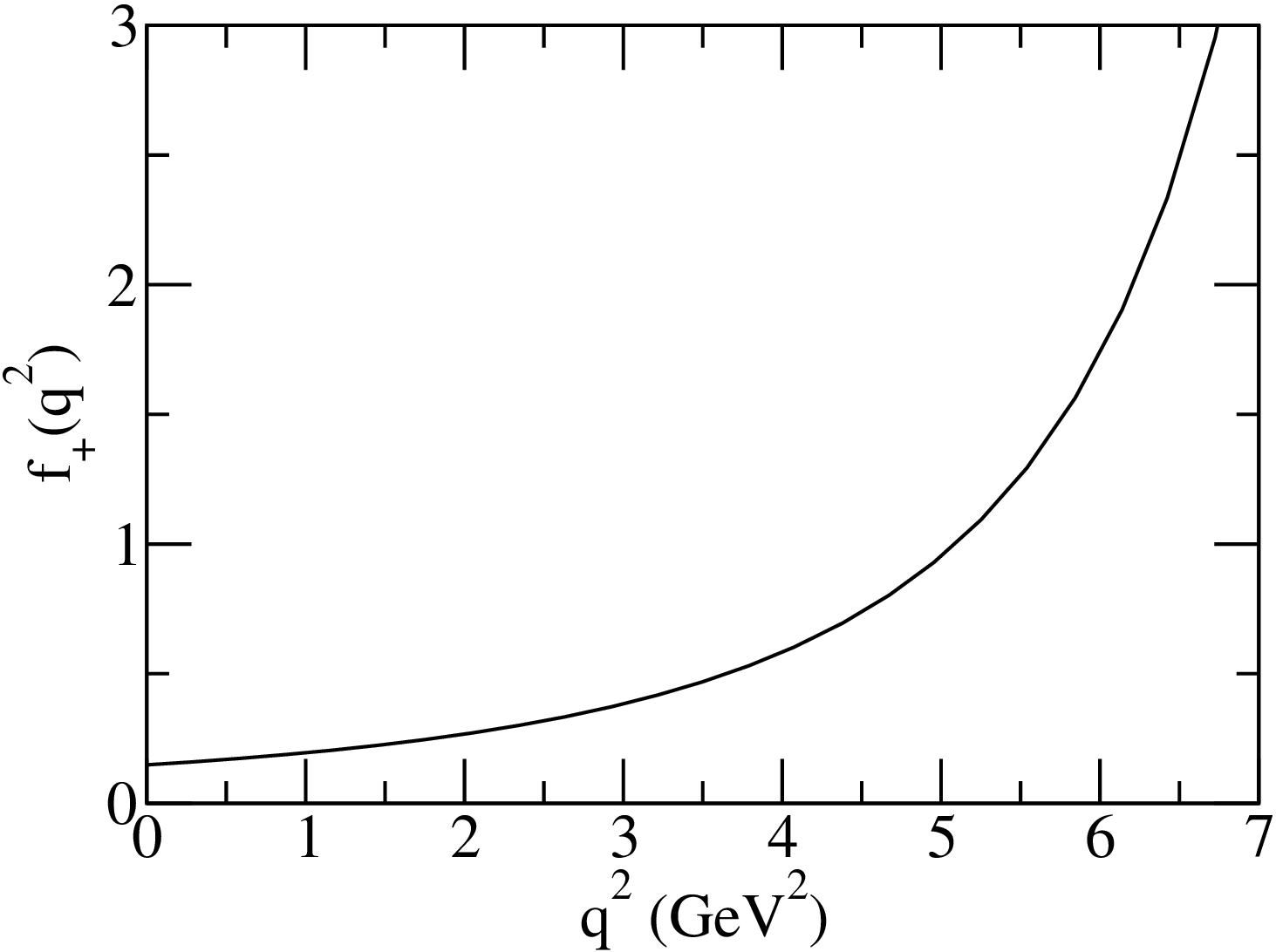}
\end{center}
\caption{The dependence of $f_{+}$ on
 $q^2$ at $M_{1}^2=15~GeV^2$, $M_{2}^2=6~GeV^2$, $s_{0}=35~GeV^2$ and $s_{0}'=9~GeV^2$.} \label{fig3}
\end{figure}

\begin{figure}
\begin{center}
\includegraphics[width=10cm]{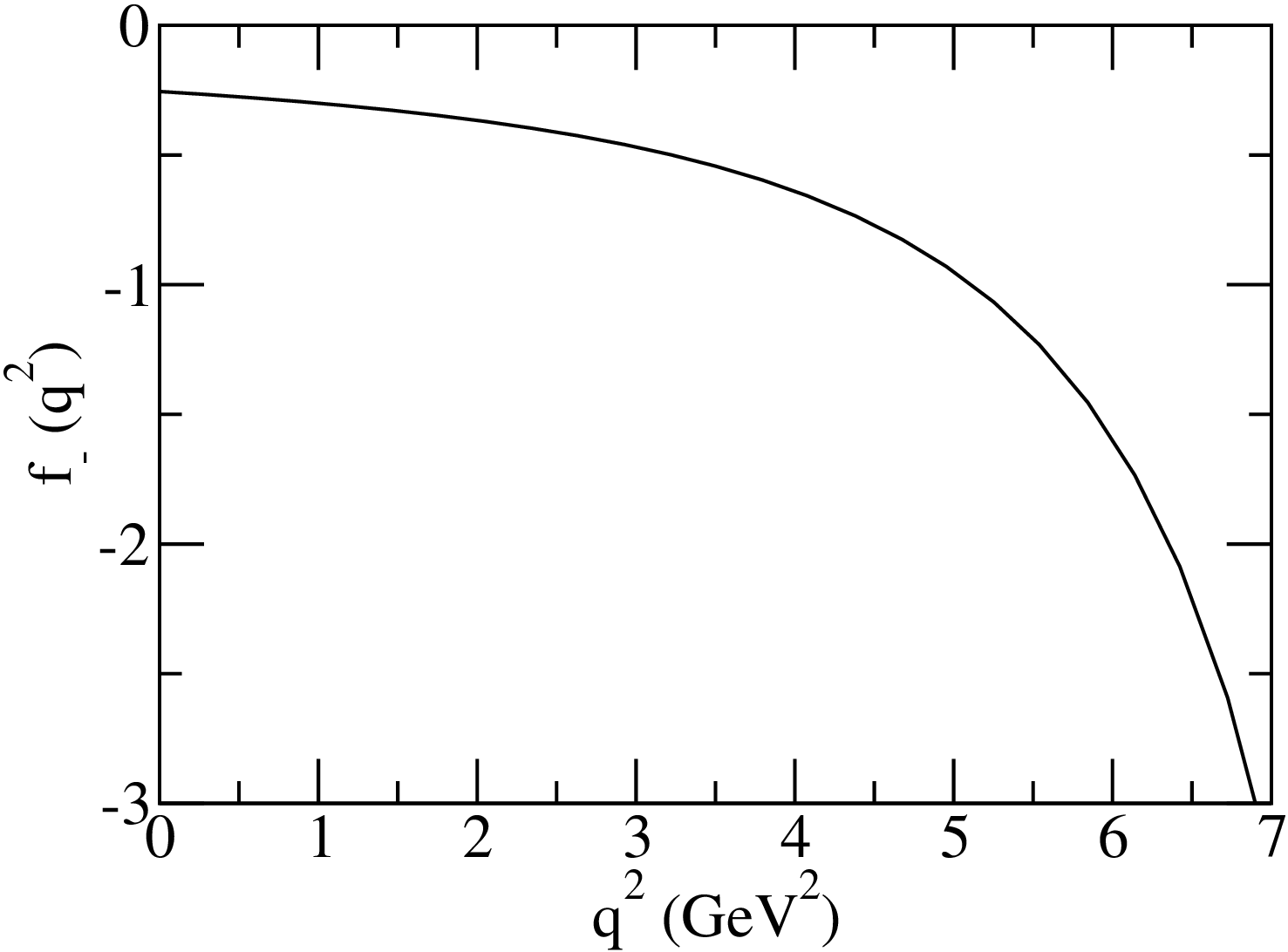}
\end{center}
\caption{The dependence of $f_{-}$ on
 $q^2$ at $M_{1}^2=15~GeV^2$, $M_{2}^2=6~GeV^2$, $s_{0}=35~GeV^2$ and $s_{0}'=9~GeV^2$.} \label{fig4}
\end{figure}

 In order to estimate the width of $B_{s} \rightarrow D_{sJ}l\nu$ it is necessary to know
 the $q^2$ dependence of the form factors $ f_{V}(q^2)$, $f_{0}(q^2)$, $f_{+}(q^2)$ and $f_{-}(q^2)$ in the whole
physical region $ m_{l}^2 \leq q^2 \leq (m_{B_{s}} - m_{D_{sJ}})^2$.
The $q^2 $ dependence of the form factors can be calculated from QCD
sum rules (for details, see  \cite{15,16}). For extracting the $q^2$
dependence of the form factors from QCD sum rules we should consider
a range $ q^2$ where the correlator function can reliably be
calculated. For this purpose we have to stay approximately $1~
GeV^2$ below the perturbative cut, i.e., up to $q^2 =8 ~GeV^2$. In
order to extend our results to the full physical region, we look for
parameterization of the form factors in such a way that in the
region $0 \leq q^2 \leq 8~ GeV^2$, this parameterization coincides
with the sum rules prediction. The dependence of form factors
$f_{V}(q^2)$, $f_{0}(q^2)$, $f_{+}(q^2)$ and $f_{-}(q^2)$  on $q^2$
are given in Figs.\ref{fig1}, \ref{fig2}, \ref{fig3} and \ref{fig4},
respectively. Our numerical calculations shows that the best
parameterization of the
form factors with respect to $q^2$ are as follows:\\
 \begin{equation}\label{17au}
 f_{i}(q^2)=\frac{f_{i}(0)}{1+\tilde \alpha\hat{q}+\tilde \beta\hat{q}^2+\tilde \gamma\hat{q}^3+\tilde \lambda\hat{q}^4}
\end{equation}
where $\hat{q}=q^2/m_{B_{s}}^2$. The values of the parameters
 $f_{i}(0),\tilde \alpha,\tilde \beta,\tilde \gamma$, and $\tilde \lambda$ are
given in the Table 1.

\begin{table}[h]
\centering
\begin{tabular}{|c|c|c|c|c|c|} \hline
  & f(0)  & $\tilde \alpha$ & $\tilde \beta$& $\tilde \gamma$& $\tilde \lambda$\\\cline{1-6}
 $f_{V}$ & 1.18 & -1.87 & -1.88& -2.41& 3.34\\\cline{1-6}
 $f_{0}$ & 0.076  & 1.85 & 0.89& 19.0& -79.3\\\cline{1-6}
 $f_{+}$ & 0.13  & -7.14 & 11.6& 21.3& -59.8\\\cline{1-6}
 $f_{-}$ & -0.26  & -4.11 & -3.27& 15.2& 18.6\\\cline{1-6}
 \end{tabular}
 \vspace{0.8cm}
\caption{Parameters appearing in the form factors of the
$B_{S}\rightarrow D_{sJ}(2460)\ell\nu$}decay in a four-parameter
fit, for $M_{1}^2=15~GeV^2$, $M_{2}^2=6~GeV^2$ \label{tab:1}
\end{table}

\begin{figure}[p]
\begin{center}
\includegraphics[width=12cm]{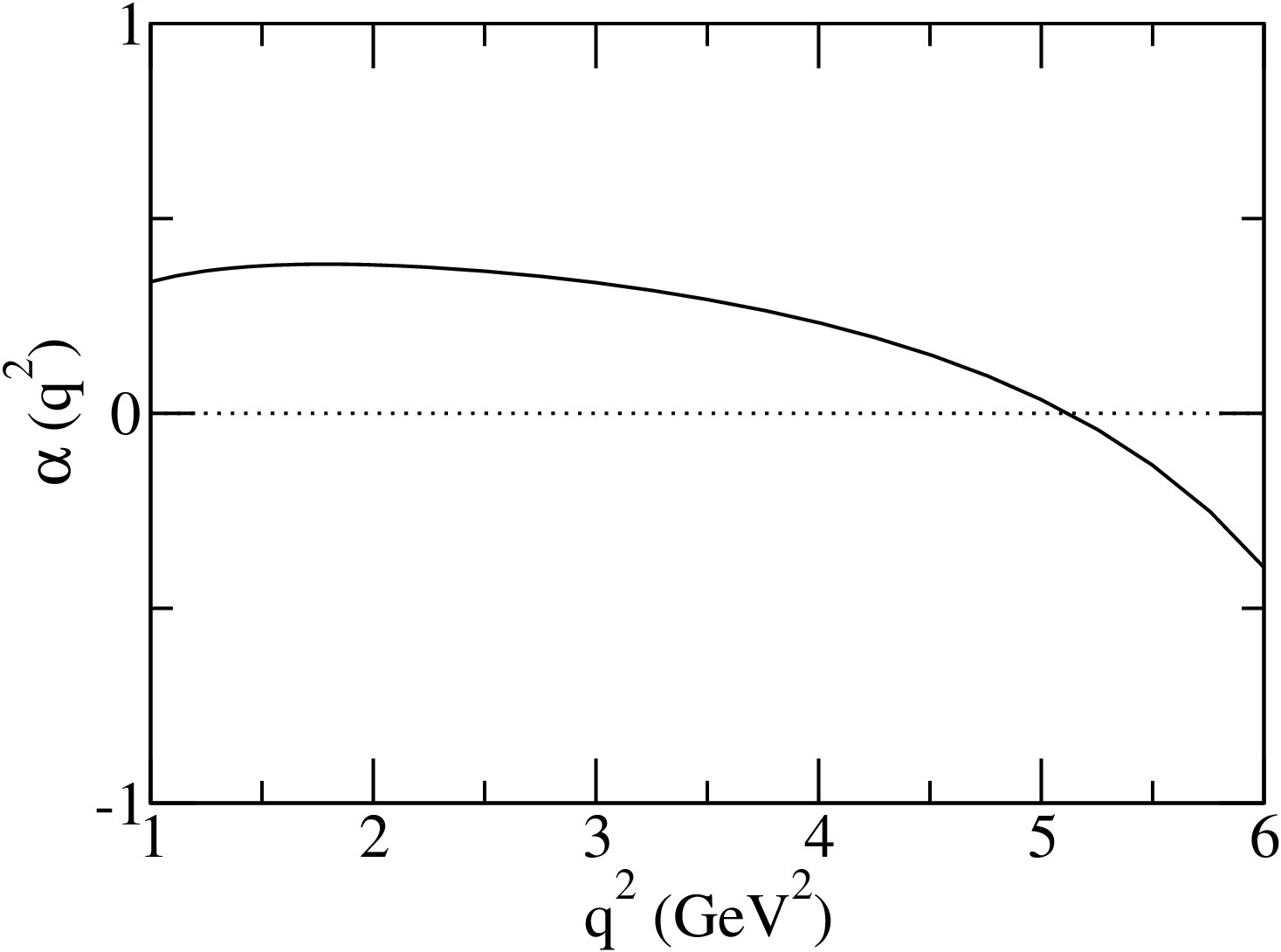}
\end{center}
\caption{The dependence of $\alpha$ on
 $q^2$.} \label{fig5}
\end{figure}

 For
$B_{s}\rightarrow D_{sJ}(2460)l\nu$ decay it is also possible to
determine the polarization of the $D_{sJ}(2460)$ meson. For this aim
we determine the asymmetry parameter $\alpha$, characterizing the
polarization of the $D_{sJ}(2460)$ meson,as
\begin{equation}\label{27au}
\alpha=2\frac{d\Gamma_{L}/dq^2}{d\Gamma_{T}/dq^2}-1
\end{equation}
 where $d\Gamma_{L}/dq^2$ and $d\Gamma_{T}/dq^2$ are differential  widths of the
decay to the states with longitudinal and transversal polarized
$D_{sJ}(2460)$ meson. After some calculations for differential decay
rates $d\Gamma_{L}/dq^2$ and $ d\Gamma_{T}/dq^2$ we get
\begin{eqnarray}\label{28au}
\frac{d\Gamma_{T}}{dq^2}=\frac{1}{8\pi^4m_{B_{s}^2}}\mid\overrightarrow{p'}\mid
G_{F}^2\mid V_{cb}\mid^2\{(2A+Bq^2)[\mid
f'_{V}\mid^2(4m_{B_{s}}^2\mid\overrightarrow{p'}\mid^2)+\mid
f'_{0}\mid^2]\}
\end{eqnarray}
\begin{eqnarray}\label{29au}
\frac{d\Gamma_{L}}{dq^2}&=&\frac{1}{16\pi^4m_{B_{s}^2}}|\overrightarrow{p'}|
G_{F}^2|V_{cb}|^2\left\{(2A+Bq^2)\left[\mid
f'_{V}\mid^2(4m_{B_{s}}^2\mid\overrightarrow{p'}\mid^2 \right.
\right. \nonumber \\ &&
+m_{B_{s}}^2\frac{\mid\overrightarrow{p'}\mid^2}{m_{D_{sJ}}}
(m_{B_{s}}^2-m_{D_{sJ}}^2-q^2))+\mid f'_{0}\mid^2 \nonumber \\ &&
-\mid
f'_{+}\mid^2\frac{m_{B_{S}}^2\mid\overrightarrow{p'}\mid^2}{m_{D_{sJ}}^2}(2m_{B_{S}}^2+2m_{D_{sJ}}^2
-q^2)-\mid
f'_{-}\mid^2\frac{m_{B_{S}}^2\mid\overrightarrow{p'}\mid^2}{m_{D_{sJ}}^2}q^2
\nonumber\\&& -2 \left.
\frac{m_{B_{S}}^2\mid\overrightarrow{p'}\mid^2}{m_{D_{sJ}}^2}(Re(f'_{0}
f_{+}^{'\ast}+f'_{0}
f_{-}^{'\ast}+(m_{B_{s}}^2-m_{D_{sJ}}^2)f'_{+}f_{-}^{'\ast}))\right]
\nonumber \\ &&
-2B\frac{m_{B_{S}}^2\mid\overrightarrow{p'}\mid^2}{m_{D_{sJ}}^2}
\left[\mid f'_{0}\mid^2+(m_{B_{s}}^2-m_{D_{sJ}}^2)^2\mid
f'_{+}\mid^2+q^4\mid f'_{-}\mid^2 \right. \nonumber \\ && +2(\left.
\left.m_{B_{s}}^2-m_{D_{sJ}}^2)Re(f'_{0}f_{+}^{'\ast})
+2q^2f'_{0}f_{-}^{'\ast}+2q^2(m_{B_{s}}^2-m_{D_{sJ}}^2)Re(f'_{+}f_{-}^{'\ast})
\right]\right\}
\nonumber \\
\end{eqnarray}
where\\
\begin{eqnarray}\label{30au}
\mid\overrightarrow{p'}\mid&=&\frac{\lambda^{1/2}(m_{B_{S}}^2,m_{D_{sJ}}^2,q^2)}{2m_{B_{S}}}\nonumber\\
A&=&\frac{1}{12q^2}(q^2-m_{l}^2)^2I_{0}\nonumber\\
B&=&\frac{1}{6q^4}(q^2-m_{l}^2)(q^2+2m_{l}^2)I_{0}\nonumber\\
I_{0}&=&\frac{\pi}{2}(1-\frac{m_{l}^2}{q^2})\nonumber\\
\end{eqnarray}
The dependence of the asymmetry parameter $\alpha$ on $q^2$ is shown
in Fig. \ref{fig5}. From this figure we see that asymmetry parameter
 $\alpha$ varies between -0.3 and 0.3 when $q^2$ lies in the region $m_{l}^2\leq q^2\leq 6~ GeV^2$.
 An interesting observation is that around $q^2=5.2 ~GeV^2$ the asymmetry parameter
changes sign. Therefore measurement of the polarization asymmetry
parameter
 $\alpha$ at fixed values of $q^2$ and determination of its
sign can give unambiguous information about quark structure of
$D_{sJ}$ meson.

 At the end of this section we would like to
present the value of the branching ratio of this decay.
 Taking into account the $q^2$ dependence of
the form factors and performing integration over $q^2$ in the limit
$m_{l}^2\leq q^2\leq(m_{B_{s}}-m_{D_{sJ}})^2$ and using the total
life-time $\tau_{B_{s}}=1.46\times10^{-12}s$ \cite{26} we get for
the branching ratio
\begin{eqnarray}\label{31au}
\textbf{\emph{B}}(B_s\rightarrow
D_{sJ}(2460)\ell\nu)\simeq4.9\times10^{-3}
\end{eqnarray}
which can be easily measurable at LHC.

In conclusion,the semileptonic
   $B_{s}\rightarrow D_{sJ}(2460)\ell\nu$ decay is
  investigated in QCD sum rule method. The $q^2$ dependence of the
  transition form factors are evaluated. The dependence of the
  asymmetry parameter $\alpha$ on $q^2$ is investigated and
  the branching ratio is estimated to be measurably large at LHC.
  
\clearpage
\newpage

\newpage
\begin{thebibliography}{99}
\bibitem{0}  T. M. Aliev, K. Azizi, A. Ozpineci,
 Eur. Phys. J. {\bf C 51} 593 (2007).
\bibitem{1}  B. Aubert et. al, BaBar,
Collaboration, Phys. Rev. Lett. {\bf 90}, 242001(2003).
\bibitem{2}  D. Besson et. al, CLEO Collaboration, Phys. Rev. {\bf D68}, 032002(2003).
\bibitem{3} Y.Nikami et. al. Belle Collaboration, Phys. Rev.
Lett. 92, 012002(2004).
\bibitem{4}  P. Krokovny et. al, Belle Collaboration, Phys. Rev. Lett. {\bf 91},
262002(2003).
\bibitem{5} A. Drutskoy et. al. Belle Collaboration, Phys. Rev. Lett. 96,
061802(2005).
\bibitem{6} B. Aubut et. al. Babar Collaboration, Phys. Rev. Lett. 93,
181807(2004).
\bibitem{7} B. Aubut et. al. Babar Collaboration, Phys. Rev. D. 69,
031101(2004).
\bibitem{8} B. Aubut et. al. Babar Collaboration, hep-ex/0408067
\bibitem{9} P.Colangelo, F. De Fazio and R. Ferrandes, Mod. Phys.
Lett. A 19, 2083(2004)
\bibitem{10}  E. S. Swanson, Phys. Rept. {\bf 429}, 243(2006).
\bibitem {11} P. Colangelo, F. De Fazio, and
A. Ozpineci, Phys. Rev. {\bf D72}, 074004 (2005).
\bibitem{12} M. A. Shifman, A. I. Vainshtein, and V. I. Zakharov, Nucl. Phys.
{\bf B147}, 385 (1979).
\bibitem {13}  P. Colangelo and A.
Khodjamirian, in At the Frontier of Particle Physics/Handbook of
QCD, edited by M. Shifman (World Scientific, Singapore, 2001), Vol.
3, p. 1495.
\bibitem{14}  T. M. Aliev, V. L. Eletsky, and Ya. I.
Kogan, Sov. J. Nucl. Phys. {\bf 40}, 527 (1984).
\bibitem{15}   P. Ball,
V. M. Braun, and H. G. Dosch, Phys. Rev. {\bf D44}, 3567 (1991).
\bibitem{16}  P. Ball, Phys. Rev. D 48, 3190 (1993).
\bibitem{17}  A. A. Ovchinnikov and V. A. Slobodenyuk, Z. Phys. {\bf C44}, 433 (1989); V. N.
Baier and A. Grozin, Z. Phys. {\bf C47}, 669 (1990).
 \bibitem{18}  A. A.
Ovchinnikov, Sov. J. Nucl. Phys. {\bf 50}, 519 (1989).
\bibitem{19}  T. M. Aliev, M. Savci, Phys. Rev. {\bf D73}, 114070(2006)
\bibitem{20} B. L. Ioffe, Prog. Part. Nucl. Phys.
{\bf 56}, 232 (2006).
\bibitem {21} Ming Qiu Huang, Phys. Rev. {\bf D69}, 114015 (2004).
\bibitem {22} M. Neubert, Phys. Rep. {\bf 245}, 259 (1994).
\bibitem{23} T. Huang and C.W. Luo, Phys. Rev. {\bf D50}, 5775 (1994).
\bibitem{24} Y. B. Dai, C. S. Huang, C. Liu, and S. L. Zhu, Phys.
Rev. {\bf D68}, 114011 (2003).
\bibitem {25}  Adam K. Leibovich, Zoltan Ligeti, Iain W. Stewart, Mark B.
Wise Phys.Rev.Lett.  {\bf 78}  3995-3998 (1997)
\bibitem {26} S. Eidelman et. al. (Particle
Data Group), Phys. Lett. {\bf B592}, 1 (2004).
\end{thebibliography}
\end{document}